\documentclass[a4paper]{article}

\usepackage{INTERSPEECH2018}
\usepackage{epsfig,amssymb,amsmath,subfigure,epsfig,siunitx,color,multirow,graphicx,bm,textcomp,enumitem}
\usepackage{bigstrut}
\usepackage[colorlinks=true,urlcolor=blue, citecolor=black, linkcolor=black]{hyperref}
\usepackage[stable]{footmisc}

\title{Waveform to Single Sinusoid Regression to Estimate the F0 Contour from Noisy Speech Using Recurrent Deep Neural Networks}
\name{Akihiro Kato, Tomi Kinnunen}
\address{University of Eastern Finland}
\email{akihiro.kato@uef.fi, tomi.kinnunen@uef.fi}

\begin{document}

\maketitle
\begin{abstract}
The fundamental frequency ($F0$) represents pitch in speech that determines prosodic characteristics of speech and is needed in various tasks for speech analysis and synthesis. Despite decades of research on this topic, $F0$ estimation at low signal-to-noise ratios (SNRs) in unexpected noise conditions remains difficult. This work proposes a new approach to noise robust $F0$ estimation using a recurrent neural network (RNN) trained in a supervised manner. Recent studies employ deep neural networks (DNNs) for $F0$ tracking as a frame-by-frame classification task into quantised frequency states but we propose \emph{waveform-to-sinusoid regression} instead to achieve both noise robustness and accurate estimation with increased frequency resolution.

Experimental results with \emph{PTDB-TUG} corpus contaminated by additive noise (\emph{NOISEX-92}) demonstrate that the proposed method improves gross pitch error (GPE) rate and fine pitch error (FPE) by more than 35 \% at SNRs between -10 dB and +10 dB compared with well-known noise robust $F0$ tracker, PEFAC. Furthermore, the proposed method also outperforms state-of-the-art DNN-based approaches by more than 15 \% in terms of both FPE and GPE rate over the preceding SNR range.
\end{abstract}

\section{Introduction\label{sec:intro}}
\emph{Fundamental frequency} ($F0$) is the lowest frequency in a quasi-periodic signal. It represents pitch in speech that determines prosodic characteristics of speech. Therefore, $F0$ is one of the key features of speech and $F0$ estimation is vital for many applications, e.g.\ voice conversion \cite{mohammadi17}, speaker and language identification \cite{torres17, nandi17}, prosody analysis \cite{godoy17}, speech coding \cite{rajendran17}, speech synthesis \cite{wang17-2} and speech enhancement \cite{kato14, kato16}.

Over the past decades, various approaches to $F0$ estimation have been proposed. Specifically, \emph{robust algorithm for pitch tracking} (RAPT) \cite{talkin95} and YIN \cite{kawahara02} that track $F0$ from time-domain signals have been widely used in many applications showing high accuracy \cite{pirker11}. These methods, however, do not attain satisfactory performance under noisy conditions \cite{wang14}. Thus, several more noise robust methods have been proposed. For instance, \emph{pitch estimation filter with amplitude compression} (PEFAC) \cite{gonzalez14} tends to outperform both RAPT and YIN in terms of noise robustness. It analyses noisy signals in the log-frequency domain with a matched filter and normalisation with the universal long-term average speech spectrum. Nonetheless, it remains challenging to obtain satisfactory estimates of $F0$ at low signal-to-noise ratios (SNRs) such as 0 dB and below.

In addition to such real-time digital signal processing (DSP) methods, various \emph{machine learning} approaches using \emph{Gaussian mixture models} (GMMs) and \emph{hidden Markov models} (HMMs) \cite{milner07, jin11}, for example, have been developed for noise robust $F0$ estimation. Furthermore, recent research has successfully applied \emph{deep neural networks} (DNNs) and their variants, e.g.\ \emph{convolutional neural networks} (CNNs) and \emph{recurrent neural networks} (RNNs), to improve $F0$ estimation in severe noise conditions \cite{wang17-2, han14-2, wang17-1}. DNNs derive discriminative models to represent arbitrarily complex mapping functions as long as they comprise enough number of units in their hidden layers. Consequently, they enable statistical models to deal with higher dimensional input features having stronger correlation than the preceding approaches.

Recently, another technical trend in acoustic modelling has emerged since a remarkable achievement of \emph{WaveNet} \cite{oord16}, which analyses time-domain waveforms directly instead of extracting spectral or cepstral features from speech. This contributed to not only advancement in speech synthesis but also end-to-end modelling for various speech applications that do not require traditional Fourier analysis \cite{oord16}. Direct analysis of waveforms is also beneficial for denoising of speech that usually combines noisy phase spectra with enhanced magnitude spectra to reconstruct clean speech \cite{rethage17}.

In fact, the latest research has applied direct time-domain waveform analysis to $F0$ estimation with DNN-based \cite{verma16} and CNN-based \cite{kim18} approaches showing improved noise robustness over both the conventional real-time signal processing and the recent DNN-based spectral analysis. These state-of-the-art time-domain $F0$ estimators, however, still have a problem to be solved: they employ DNNs or CNNs to form a frame-by-frame \emph{classification} model to decide a state corresponding to a \emph{quantised} frequency. Even if it is convenient to treat $F0$ tracking as a classification task in the same manner as alignment of senones in speech recognition, the resultant estimates of $F0$ contours have a limited frequency resolution determined by the number of quantised frequency states. This is a potential draw-back in terms of estimation accuracy of $F0$.

This work is an extension of our recent preliminary study \cite{kato18-1}. In that study, we have successfully employed an RNN regression model, which maps spectral sequence directly onto $F0$ values, to tackle the disadvantage in existing classification approaches mentioned above. In relation to that preliminary study, the present paper represents the following four major changes. First, we employ direct waveform inputs instead of spectral sequence. Second, we propose a novel encoding method of the $F0$ information using a simple sinusoid oscillated with the ground truth value of $F0$. This encoding enables our model to map raw speech waveforms to raw sinusoids without requirement of neither pre-processing nor post-processing. Next, we amend our experiments with very recent competitive methods which are also based on waveform input schemes \cite{verma16, kim18}. Finally, we augmented noise conditions for the experiments in order to examine noise robustness against more various noise types. Consequently, a known noise condition is increased from consisting of six noise types to eight types while an unknown noise condition is augmented from two types to four types.

\section{Methodology\label{sec:method}}
\begin{figure}[htbp]
	\begin{center}
		\includegraphics[scale=0.38]{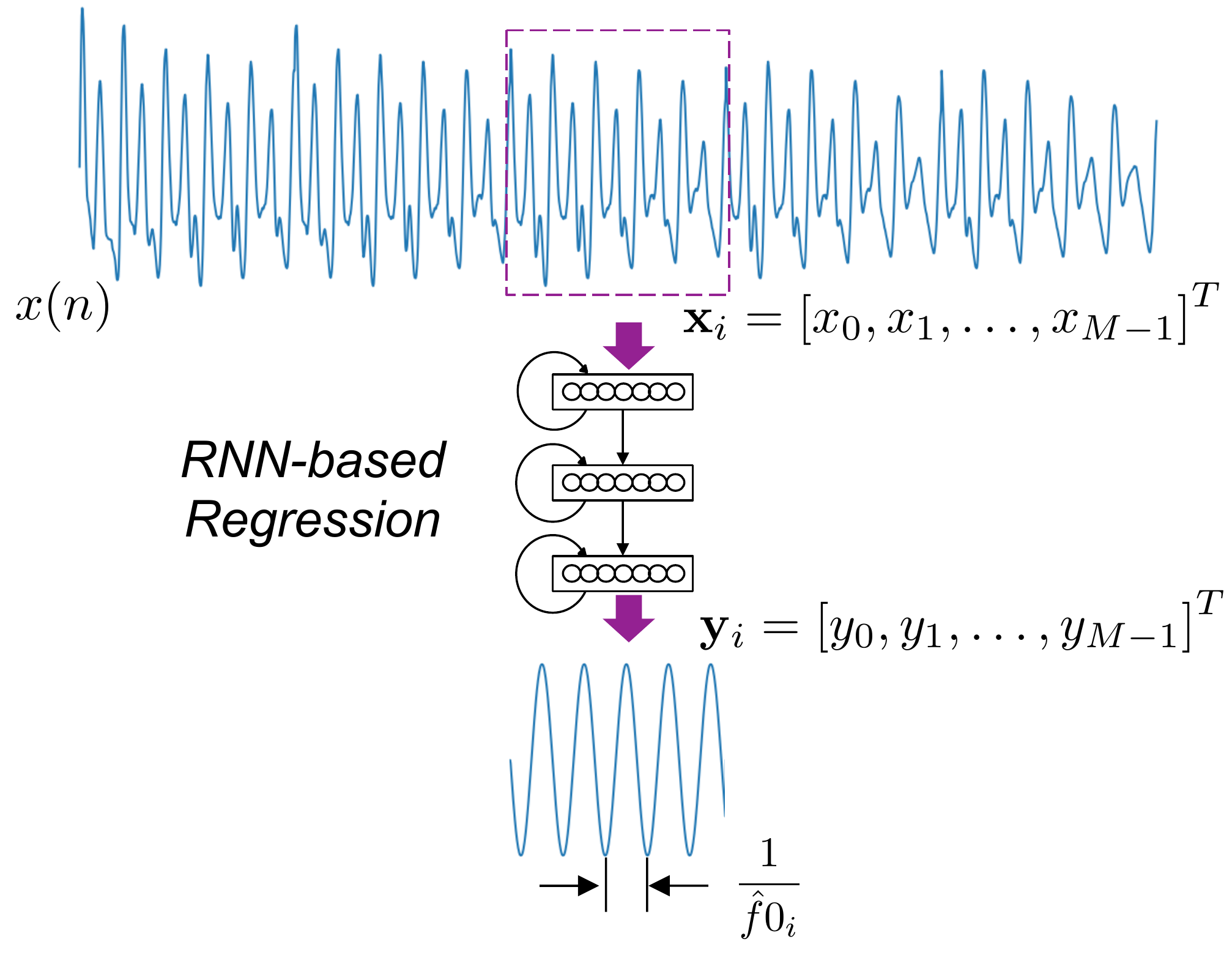}
		\caption{\it A voiced speech waveform directly inputs to an RNN in order to perform waveform-to-sinusoid regression. The estimate of $F0$ is then inferred from the resultant sinusoid.}
		\label{fig:framework}
	\end{center}
\end{figure}

In the proposed method of $F0$ estimation, discrete time-domain speech waveform, $x(n)$, is used as an input to an RNN. For voiced input speech, the posteriors of the RNN are mapped onto a single sinusoid oscillated with $F0$ of the input waveform as a regression task. For unvoiced and no voice inputs, the RNN performs an identity mapping. The $F0$ value is then explicitly inferred from the resultant single sinusoid using its autocorrelation. Figure \ref{fig:framework} illustrates the proposed framework.

\subsection{Waveform-to-sinusoid regression using an RNN}
$x(n)$ is first divided into $I$ frames, ${\bf x}_0$, ${\bf x}_1$, \dots, ${\bf x}_{I-1}$, 
\begin{equation}
	{\bf x}_i = \left[x(Mi), x(Mi+1), \dots, x(M(i+1)-1)\right]^\top,
\end{equation}
where $M$ denotes the number of samples in a frame.

Units in RNN layers have connections from their outputs back to their own inputs in addition to the feedforward connections. Therefore, an RNN layer receives its own output at the previous time sequence as well as the current time sequence input from the previous layer. This behaviour of RNN layers, interpreted as memory cells \cite{grossberg75}, is well-suited to analyse temporal dynamics of speech. Thus, $F0$ at the $i$-th frame, $f0_i$, is analysed with neighbouring $2p$ frames, i.e.\ $i-p$ to $i+p$, sequence-by-sequence.

Since the RNN-based discriminative model in the proposed method takes a form of sequence-to-sequence structure, output of RNN layer, $l$, at time sequence, $n$ $(n = 0, 1, \dots, 2p)$, ${\bm \theta}_n^l$, is derived as follows with respect to input instance, ${\bf x}_i$.
\begin{eqnarray}
	{\bm \theta}_n^l &=& g\left({\bf W}^l{\bm \phi}_n^l + {\bf H}^l{\bm \phi}_{n-1}^{l+1}\right)\\
	{\bm \phi}_n^l &=& \left[1,~ ({\bm \theta}_n^{l-1})^\top\right]^\top\\
	{\bm \theta}_n^0 &=& \left[1,~ ({\bf x}_{i-p+n})^\top\right]^\top,
\end{eqnarray}
where $g(\cdot)$ represents an activation function and ${\bf W}^l$ is the weight matrix from the output of layer $l-1$ to the input of layer $l$ (feedforward) while ${\bf H}^l$ denotes the weight matrix from the output of layer $l$ to the input of layer $l$ (feedback). Here, ${\bf W}^l$ and ${\bf H}^l$ are represented as
\begin{eqnarray}
	{\bf W}^l &=& \left[
	\begin{array}{cccc}
		w_{10}^l & w_{11}^l & \dots & w_{1q_{l-1}}^l\\
		w_{20}^l & w_{21}^l & \dots & w_{2q_{l-1}}^l\\
		\vdots & \vdots & \ddots & \vdots\\
		w_{q_l0}^l & w_{q_l1}^l & \dots & w_{q_lq_{l-1}}^l
	\end{array}
	\right]\\
	{\bf H}^l &=& \left[
	\begin{array}{cccc}
		h_{10}^l & h_{11}^l & \dots & h_{1q_l}^l\\
		h_{20}^l & h_{21}^l & \dots & h_{2q_l}^l\\
		\vdots & \vdots & \ddots & \vdots\\
		h_{q_l0}^l & h_{q_l1}^l & \dots & h_{q_lq_l}^l,
	\end{array}
	\right],
\end{eqnarray}
where $q_l$ is the number of units (excluding the bias unit) in the $l$-th layer. Furthermore, $w_{jk}^l$ is the feedforward weight between unit, $k$, in the ($l-1$)-th layer and unit, $j$, in the $l$-th layer while $h_{jk}^l$ is the feedback weight between the output of unit, $k$, and the input of unit, $j$, in the $l$-th layer.

In order to achieve the RNN-based regression to a single sinusoid, the output layer is activated by the identity function unlike classification model applying the softmax function for the activation. Consequently, posteriors of the RNN at the $i$-th frame (i.e.\ $n = p$), ${\bf y}_i$, is derived as
\begin{eqnarray}
	{\bf y}_i &=& \mathcal{I}\left({\bf W}^L{\bm \phi}_p^L + {\bf H}^L{\bm \phi}_{p-1}^{L+1}\right)\\
	{\bm \phi}_p^{L+1} &=& \left[1,~ {\bf y}_{p-1}^\top\right]^\top,
\end{eqnarray}
where $\mathcal{I}(\cdot)$ and $L$ denote the identity function and the number of RNN layers respectively.

\subsection{Model training and F0 estimation}
\begin{figure}[htbp]
	\begin{center}
		\includegraphics[scale=0.4]{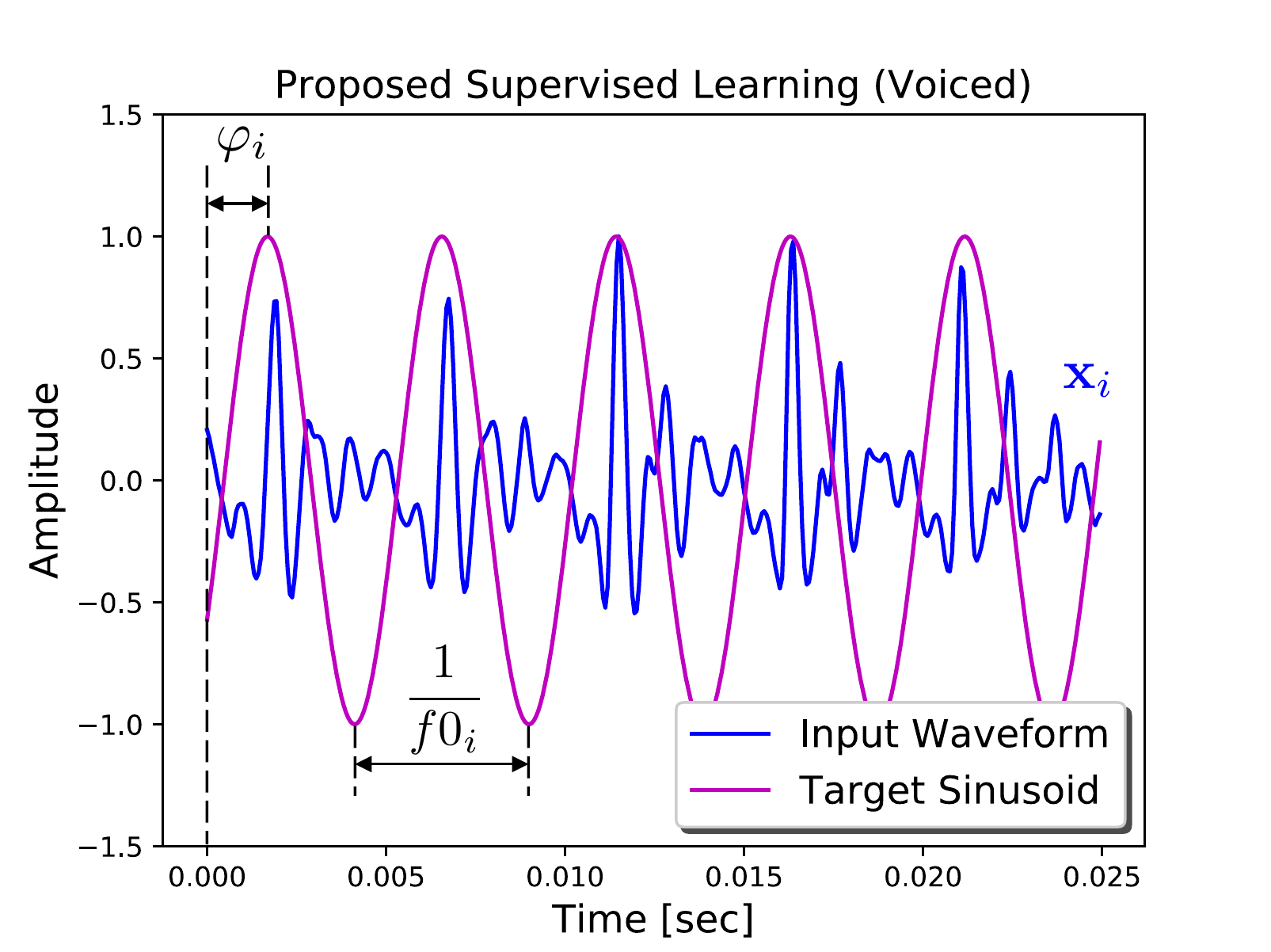}
		\caption{The target sinusoid for supervised training is oscillated with the ground truth of $F0$ at frame, $i$ ($f0_i$).}
		\label{fig:sinusoid}
	\end{center}
\end{figure}

In the offline training process, ${\bf W}^l$ and ${\bf H}^l$ are optimised in advance by supervised learning. During \emph{voiced} speech periods, training is achieved by minimising the mean square error (MSE) between ${\bf y}_i$ and its target sinusoid. The sinusoid is oscillated with $f0_i$ and $\varphi_i$ which are the ground truth of $F0$ at frame, $i$, and the phase to maximise the cross-correlation between ${\bf x}_i$ and $\cos(2\pi f0_im/f_s)$. Here, $f_s$ is the sampling frequency and $m = 0, 1, \dots, M-1$, as illustrated in Figure \ref{fig:sinusoid}.
For \emph{unvoiced} and \emph{no voice} periods, the target of supervised learning is set to the input waveform itself, i.e.\ identity mapping. The weight optimisation during this training is accomplished with mini-batch gradient descent with the backpropagation algorithm \cite{rumelhart85}.

\begin{figure}[htbp]
	\begin{center}
		\includegraphics[scale=0.42]{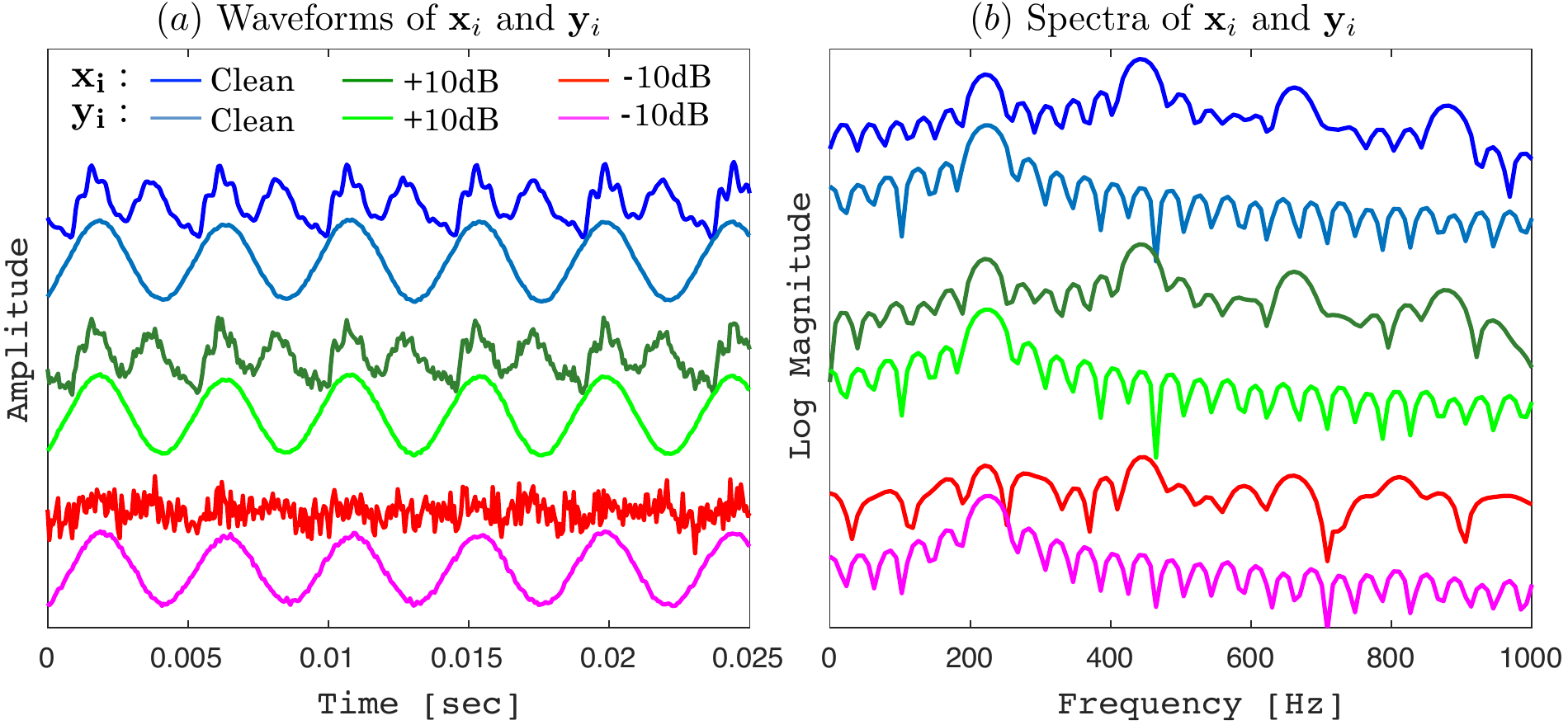}
		\caption{\it (a) depicts the relation between ${\bf x}_i$ and ${\bf y}_i$ in clean and noisy conditions (SNRs at +10 dB and -10 dB in white noise). (b) plots their magnitude spectra. Waveforms and spectra in each plot have been shifted for better visualisation.}
		\label{fig:posterior}
	\end{center}
\end{figure}

Finally, estimate of $F0$ at the $i$-th frame including voiced, unvoiced and no voice frames, $\hat{f}0_i$, is inferred from ${\bf y}_i$ by maximising its autocorrelation. Since unvoiced and no voice frames are transformed with identity mapping, those frames can be effectively detected if the maximum of cross correlation between ${\bf y}_i$ and $\cos (2\pi \hat{f}0_i m/f_s)$ is lower than threshold, $\lambda$. In other words, voiced frames are easily distinguished if the shape of ${\bf y}_i$ is close to a sinusoid, otherwise the frame is unvoiced or no voice. $\lambda$ is empirically selected as 0.15 by preliminary cross-validation test. Figure \ref{fig:posterior} (a) demonstrates the relation between input voiced waveform, ${\bf x}_i$, and sinusoid obtained by the proposed RNN model, ${\bf y}_i$, at different SNRs in white noise. (b) plots their magnitude spectra to illustrate how ${\bf y}_i$ is suitable for $F0$ analysis compared with ${\bf x}_i$.


The RNN model above can be replaced with more sophisticated recurrent units such as the \emph{long short-term memory} (LSTM) cells \cite{hochreiter97} or the \emph{gated recurrent unit} (GRU) cells \cite{chung14} in order to capture longer-term dependencies in signals. We employ LSTM cells for our approach in the following experiments.

\section{Experiments\label{sec:result}}
We examine both accuracy and noise robustness of the proposed methods and compare them with \emph{RAPT} \cite{talkin95}, \emph{YIN} \cite{kawahara02}, \emph{PEFAC} \cite{gonzalez14} and state-of-the-art DNN-based classification approaches: \emph{DNN-CLS(S)} \cite{han14-2}, \emph{DNN-CLS(W)} \cite{verma16} and \emph{CREPE} \cite{kim18}. DNN-CLS(S) is based on a DNN classification model from spectral features to quantised frequency states whereas DNN-CLS(W) uses a DNN classification model from waveforms to quantised frequency states. CREPE is an $F0$ tracker targeting at music signals using a CNN-based classification model from waveforms to quantised frequency states.

Performance of these $F0$ trackers is evaluated in terms of two standard metrics: \emph{gross pitch error} (GPE) rate and \emph{fine pitch error} (FPE) \cite{rabiner76}. GPE frames represent voiced frames in which the error between the estimated pitch period ($1/\hat{f}0$) and the ground truth ($1/f0$) is more than 10 samples, i.e.\ 0.625 ms. FPE frames, in turn, are voiced frames excluding GPE frames. The mean of FPEs, $\mu_{\text{FPE}}$, represents the bias in $F0$ estimation whereas the standard deviation of FPEs, $\sigma_{\text{FPE}}$, measures the accuracy of estimation \cite{rabiner76}.

\subsection{Datasets (PTDB-TUG corpus + NOISEX-92)}
The experiments use speech from \emph{pitch tracking database from Graz University of Technology} (PTDB-TUG) \cite{pirker11}. The training set consists of 3200 utterances spoken by 16 speakers (8 males, 8 females), i.e.\ 200 utterances each. The cross-validation (CV) set comprises other 576 utterances spoken by the same 16 speakers, 36 utterances per each. For the test set, 944 utterances spoken by 4 speakers (2 males, 2 females) who are not in the training and CV sets (unknown speakers), i.e.\ 236 utterances each, are contained in order to set the test condition as \emph{speaker independent} (SI).

Speech in each dataset is sampled at 16kHz and the sampled signals in the training and CV sets are contaminated with eight types of additive noise at five levels of SNR, -10, -5, 0, +5 and +10 dB. The noise types are referred to as \emph{Babble}, \emph{Destroyerops}, \emph{F16}, \emph{Factory2}, \emph{Leopard}, \emph{M109}, \emph{Machinegun} and \emph{White} in NOISEX-92 \cite{varga93}. The test set is contaminated with other four types of additive noise including \emph{Destroyerengine}, \emph{Factory1}, \emph{Pink} and \emph{Volvo} in addition to the preceding eight types at the same SNRs as the training and CV sets. The former four types make an \emph{unknown noise} condition while the latter eight types give a \emph{known noise} condition. Consequently, the training set amounts 131,200 utterances (15,252 min), i.e.\ 3,200 $\times$ (8 noise $\times$ 5 level + 1 clean), and the CV set becomes 23,616 utterances (2,542 min) while the test set amounts 57,584 utterances (6,344 min), i.e.\ 944 $\times$ (12 noise $\times$ 5 level + 1 clean), in total. 

PTDB-TUG contains ground truth $F0$ contours of each utterance obtained from laryngograph signals recorded in a clean condition to which a Kaiser filter and RAPT were applied. These are used in the following experiments as the ground truth.

\subsection{Training and test settings}
The speech signals in the datasets are framed into 25 ms frames at 5 ms intervals. The first 400 frames and the last 200 frames of each utterance are then removed to reduce non-speech frames.

The hyperparameters of the proposed method are empirically selected by preliminary cross-validation test. The number of hidden layers (LSTM cells) are set equal to three with 1024 units each that are activated by $\tanh$ function. Mini-batch size is set to 300 frames and random unit dropout (25 \%) and batch normalisation \cite{ioffe15} are applied during training. To train or analyse frame, ${\bf x}_i$, 15 neighbouring frames in a row, i.e.\ from ${\bf x}_{i-7}$ to ${\bf x}_{i+7}$ are used as inputs to the RNN to perform sequence-to-sequence analysis.

Feature extraction and parameter settings of the other methods follow their original paper mentioned above but the posterior frequency states in CREPE are modified with the same quantisation manner as DNN-CLS(S\&W) because the classification target of original CREPE is music signals.

\subsection{Results and discussion}
\begin{figure}[htbp]
	\begin{center}
		\includegraphics[scale=0.45]{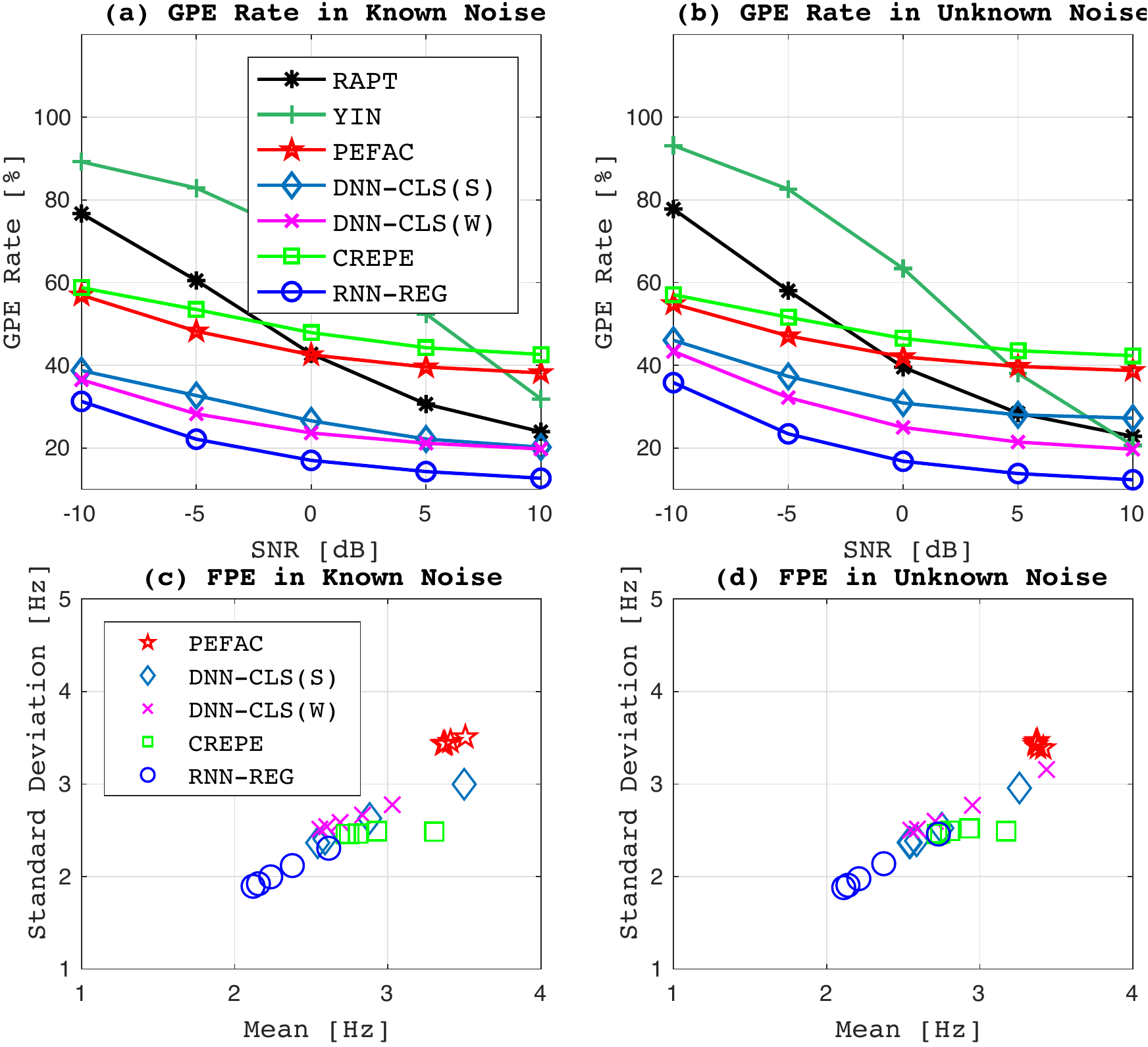}
		\caption{\it $F0$ estimation performance of each method at different SNRs showing (a) GPE rates in the known noise condition, (b) GPE rates in the unknown noise condition. (c) illustrates a scatter plot of $\mu_{\text{FPE}}$ and $sigma_{\text{FPE}}$ in the known noise condition while (d) shows the performance in the unknown noise.}
		\label{fig:result}
	\end{center}
\end{figure}

Figure \ref{fig:result} (a) illustrates GPE rates of each method at different SNRs in the multi noise condition of the known noise types. (b) represents GPE rates in the multi noise condition of the unknown noise. 
RNN-REG always shows the best performance in terms of GPE rate. It outperforms the other methods over the SNR range between -10 and +10 dB in both known and unknown noise conditions giving GPE rate of around 35 \% at -10 dB. DNN-CLS(S\&W) also show lower GPE rate than the other real-time DSP methods, i.e.\ RAPT, YIN and PEFAC, but they always exceed RNN-REG by around 8 or more percentage points in both noise conditions. CREPE is not as robust as the other three neural net-based methods in terms of GPE rate.

GPE frames correspond to failure in $F0$ estimation at voiced frames \cite{rabiner76}. In that sense, $F0$ estimation with YIN at SNRs below 5 dB, RAPT at less than 0 dB and CREPE and PEFAC at -5 dB and below are likely to have unreliable frames accounting for more than 50 \% of voiced frames. Conversely, RNN-REG keeps estimation failure approximately 35 \% of voiced frames even at -10 dB in unknown noise condition whereas DNN-CLS(S\&W) score over 40 \% at -10 dB in unknown noise. This demonstrates substantial advantage of our proposal in $F0$ estimation from noisy speech.

Figures \ref{fig:result} (c) and (d) illustrate the performance of PEFAC, DNN-CLS(S\&W), CREPE and RNN-REG in terms of FPE at SNRs of -10, -5, 0, +5 and +10 dB in the known and unknown noise conditions respectively as scatter plots of $\mu_{\text{FPE}}$ and $\sigma_{\text{FPE}}$. YIN and RAPT are eliminated from this evaluation because sufficient amount of frames for FPE analysis are not brought by those methods in such noisy conditions.

Since $\mu_{\text{FPE}}$ represents the bias in $F0$ estimation while $\sigma_{\text{FPE}}$ is a measure of the accuracy in the estimation \cite{rabiner76}, RNN-REG performs best in terms of both bias and accuracy of estimation over the SNR range between -10 dB and +10 dB in both known and unknown noise conditions. Although PEFAC shows strong noise robustness in both accuracy and bias, RNN-REG outperforms it by approximately 35 \% on average in both known and unknown noise. RNN-REG also superior to DNN-CLS(S), DNN-CLS(W) and CREPE by more than 15 \%.

In comparison among RNN-REG, DNN-CLS(S\&W) and CREPE, the regression task to map waveforms onto the sinusoid encoding $F0$ is more difficult than the classification task to classify the waveforms or spectral features into quantised frequencies. However, RNN regression can capture temporal dynamics by optimising recurrent weights unlike the full-connected DNN in DNN-CLS(S) augmenting the input with consecutive frames which produce a lot of poor-correlated connections into the network, e.g.\ a connection between a unit in a past frame and a unit in a future frame. Consequently, RNN regression accuracy outperforms the quantised frequencies in the classification task although the resolution of RNN-REG is also restricted by the sampling period.

\begin{figure}[htbp]
	\begin{center}
		\includegraphics[scale=0.45]{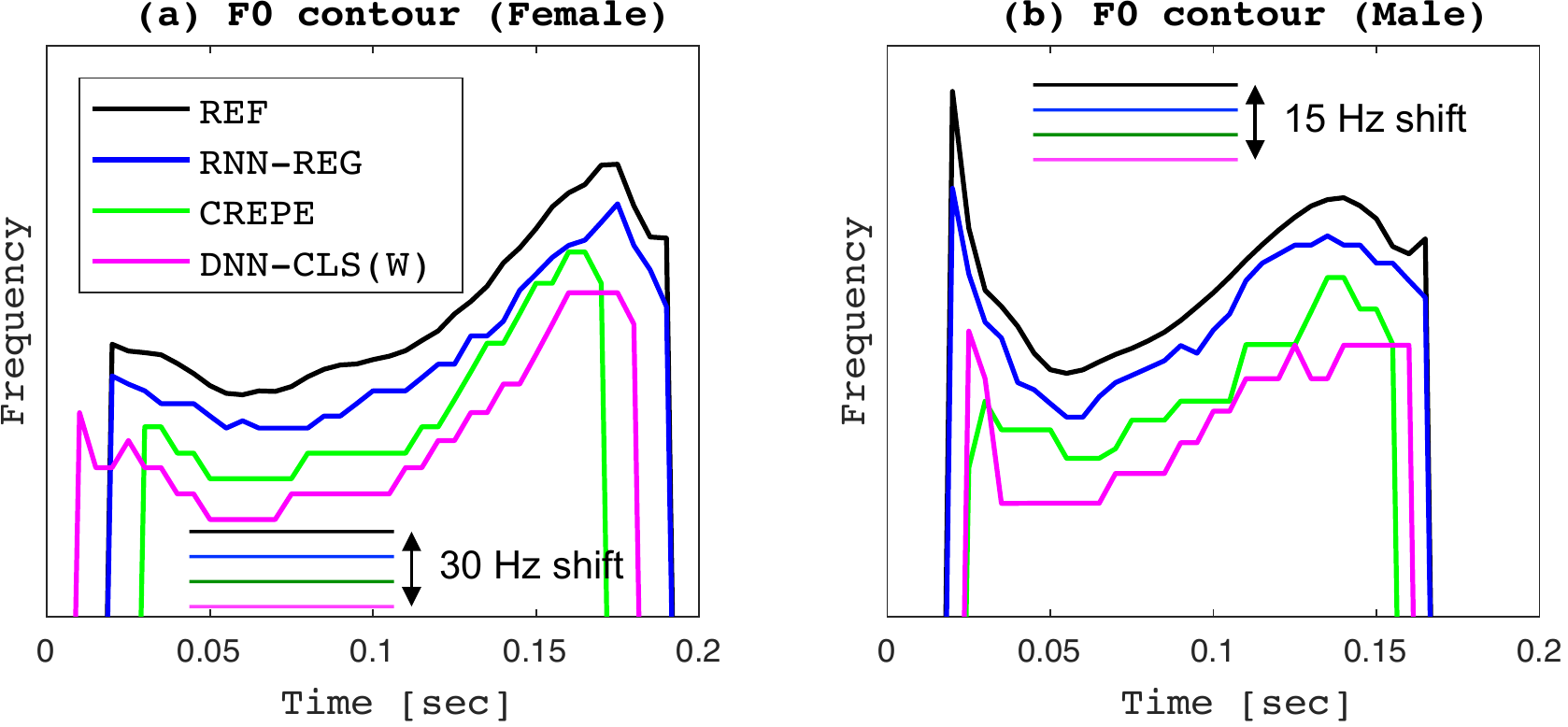}
		\caption{\it $F0$ contours of word ``DARK'' spoken by (a) an unknown female speaker and (b) an unknown male speaker. $F0$ contours in plot (a) are shifted at 10 Hz intervals while the contours in plot (b) are shifted at 5 Hz intervals for better visualisation.}
		\label{fig:contour}
	\end{center}
\end{figure}

Figure \ref{fig:contour} illustrates $F0$ contours of the spoken word ``DARK'' estimated by DNN-CLS(W), CREPE and RNN-REG in a clean condition and they are compared with the ground truth (\emph{REF}). (a) and (b) show the $F0$ contours spoken by a female speaker and a male speaker respectively. Utterances of these two speakers are not included in the training set, i.e.\ unknown speakers.
The figures demonstrate the advantage of the proposed method employing RNN-based waveform-to-sinusoid regression approach over classification approaches using DNNs or CNNs. Specifically, the $F0$ contours estimated by RNN-REG is closer to the ground truth than other methods. This also reveals the potential of our proposal (RNN-REG) to track prosody of different speakers in a speaker-independent manner.

\section{Conclusion}
We addressed the problem of $F0$ estimation with a waveform-to-sinusoid regression using an RNN in order to obtain accurate $F0$ estimates with improved noise robustness. The proposed RNN-based approach demonstrates considerable improvement over the existing state-of-the-art $F0$ trackers. Compared to PEFAC, one of the most robust autocorrelation-based $F0$ trackers, the proposed method yielded a relative improvement exceeding 35 \% in both gross pitch error (GPE) rate and fine pitch error (FPE) at SNRs between -10 dB and +10 dB in both known and unknown noise conditions. Furthermore, the proposed method outperformed the latest DNN and CNN-based $F0$ trackers, in terms of relative improvement in both GPE rate and FPE, by more than 15 \% over the preceding SNR range.

Comparison of the estimated $F0$ contours of clean speech also demonstrates an advantage of our proposal over other DNN and CNN-based approaches in producing more natural $F0$ trajectories. While the present work focused solely on the $F0$ tracking problem itself, our future plan involves integrating the proposed method in a downstream application such as voice conversion or prosody-based speaker recognition.

\section{Acknowledgement}
This work was supported in part by Academy of Finland (Proj. No. 309629). The authors wish to acknowledge CSC - IT Centre for Science, Finland, for computational resources.

\bibliographystyle{IEEEtran}
\bibliography{mybib}

\end{document}